\documentstyle[11pt,paspconf,psfig]{article}

\newcommand{\simgt}{\lower.5ex\hbox{$\; \buildrel > \over \sim \;$}}
\newcommand{\simlt}{\lower.5ex\hbox{$\; \buildrel < \over \sim \;$}}
\newcommand{\A}{{\scriptscriptstyle A}}
\newcommand{\B}{{\scriptscriptstyle B}}
\newcommand{\dL}{{d_{\scriptscriptstyle L}}}

\newcommand{\gas}{{\rm\scriptscriptstyle gas}}
\newcommand{\bol}{{\rm\scriptscriptstyle bol}}
\newcommand{\band}{{\rm\scriptscriptstyle band}}
\newcommand{\ns}{Log$N$--Log$S$~}
\newcommand{\unit}{erg cm$^{-2}$ s$^{-1}$}
\begin{document}

\title{Cosmological Implications of X-ray Clusters of Galaxies}

\author{Yasushi Suto and Tetsu Kitayama}
\affil{Department of Physics and Research Center for the Early
Universe \\ The University of Tokyo, Tokyo 113-0033, Japan}

\author{Shin Sasaki}
\affil{Department of Physics, Tokyo Metropolitan University, 
Hachioji\\ Tokyo 192-0364, Japan}

\begin{abstract}
  Cosmological implications of clusters of galaxies are discussed with
  particular attention to their importance in probing the cosmological
  parameters. More specifically we compute the number counts of
  clusters of galaxies, \ns relation, in X-ray and submm bands on the
  basis of the Press--Schechter theory. We pay particular attention to
  a set of theoretical models which well reproduce the {\it ROSAT}
  0.5-2 keV band \ns, and explore possibilities to break the
  degeneracy among the viable cosmological models.
\end{abstract}


\section{Introduction}

There are several reasons why clusters of galaxies are regarded as
useful probes of cosmology including (i) since dynamical time-scale of
clusters is comparable to the age of the universe, they should retain
the cosmological initial condition fairly faithfully.  (ii) clusters
can be observed in various bands including optical, X-ray, radio, mm
and submm bands, and in fact recent and future big projects (e.g.,
SDSS, AXAF, PLANCK) aim to make extensive surveys and detailed
imaging/spectroscopic observations of clusters.  (iii) to the first
order, clusters are well approximated as a system of dark matter, gas
and galaxies, and thus theoretically well-defined and relatively
well-understood, at least compared with galaxies themselves, and (iv)
on average one can observe a higher-z universe with clusters than with
galaxies.  In particular X-ray observations are well-suited for the
study of clusters since the X-ray emissivity is proportional to
$n_e^2$ and thus less sensitive to the projection contamination which
has been known to be a serious problem in their identifications with
the optical data.

In fact, various statistics related to the abundances of clusters has
been extensively studied to constrain theories of structure formation,
including mass function (Bahcall \& Cen 1993; Ueda, Itoh, \& Suto
1993), velocity function (Shimasaku 1993; Ueda, Shimasaku, Suginohara,
\& Suto 1994), X-ray Temperature function (hereafter XTF, Henry \&
Arnaud 1991; White, Efstathiou, \& Frenk 1993; Kitayama \& Suto 1996 ;
Viana \& Liddle 1996; Eke, Cole, \& Frenk 1996; Pen 1996).  Previous
authors have focused on cosmological implications of cluster XTF
mainly because theoretical predictions are relatively easier although
the observational data are statistically limited. In addition, the
conversion to the number density at high z becomes very sensitive to
the adopted cosmological parameters.  On the other hand, \ns which we
discuss in details below is observationally more robust (Ebeling et
al. 1997; Rosati \& Della Ceca 1997) while its theoretical prediction
is more model-dependent (Oukbir, Bartlett, \& Blanchard 1996; Kitayama
\& Suto 1997; Kitayama, Sasaki \& Suto 1998).  In this respect, both
statistics are complementary.

\begin{figure}[t]
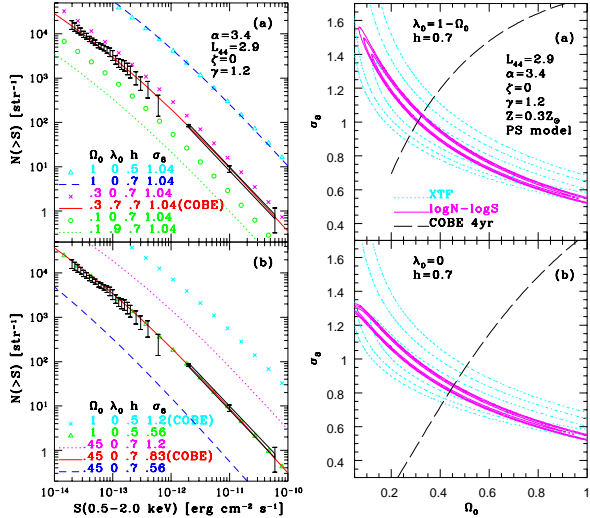

\begin{center}
\leavevmode
\psfig{figure=logns1.cps,height=6.8cm}
\psfig{figure=chi2ns2.cps,height=6.8cm}
\end{center}
\vspace*{-0.5cm}
\caption{ {\it Left:} Theoretical predictions for \ns of X-ray
  clusters in CDM models with different cosmological parameters; (a)
  $\sigma_8=1.04$ models with different $\Omega_0$, $\lambda_0$ and
  $h$, (b) $\Omega_0=1$ and $0.45$ models with different
  $\sigma_8$. Denoted by (COBE) are the models normalized according to
  the {\it COBE} 4 year data (Bunn \& White 1997).  Data points with
  error bars at $S\simlt 10^{-12}$ \unit are from the {\it ROSAT} Deep
  Cluster Survey (RDCS, Rosati et al. 1995; Rosati \& Della Ceca
  1997), and the error box at $S\simgt 2 \times 10^{-12}$ represents a
  power-law fitted region from the {\it ROSAT} Brightest Cluster
  Sample (BCS, Ebeling et al. 1997). For the BCS data at $S= 2 \times
  10^{-12}$, $1 \times 10^{-11}$ and $6 \times 10^{-11}$\unit, we also
  plot the corresponding Poisson errors.  {\it Right:} Limits on
  $\Omega_0$ and $\sigma_8$ in CDM models ($n=1$, $h=0.7$) with (a)
  $\lambda_0=1-\Omega_0$, and (b) $\lambda_0=0$. Constraints from
  cluster \ns (solid) and XTF (dotted) are plotted as contours at $1
  \sigma$(68\%), $2\sigma$(95\%) and $3\sigma$(99.7\%) confidence
  levels. Dashed lines indicate the {\it COBE} 4 year results from
  Bunn \& White (1997).  }
\label{fig:ns1chi2}
\vspace*{-0.5cm}
\end{figure}

\section{Log $N$ -- Log $S$ of X-ray clusters}

We compute the number of clusters observed per unit
solid angle with X-ray flux greater than $S$ by
\begin{eqnarray}
  N(>S)= \int_{0}^{\infty}dz ~d_\A^2(z) \, c
  \left|{\frac{dt}{dz}}\right| \int_{S}^\infty dS ~ (1+z)^3 n_M(M,z)
  \frac{dM}{dT_{\gas}}\frac{dT_{\gas}}{dL_\band} \frac{dL_\band}{dS},
\label{eq:logns}
\end{eqnarray}
where $c$ is the speed of light, $t$ is the cosmic time, $d_\A$ is the
angular diameter distance, $T_{\gas}$ and $L_{\band}$ are respectively
the gas temperature and the band-limited absolute luminosity of
clusters, and $n_M(M,z)dM$ is the comoving number density of
virialized clusters of mass $M \sim M+dM$ at redshift $z$.

Given the observed flux $S$ in an X-ray energy band [$E_a$,$E_b$], the
source luminosity $L_\band$ at $z$ in the corresponding band
[$E_a(1+z)$,$E_b(1+z)$] is written as
\begin{equation}
  L_\band[E_a(1+z),E_b(1+z)] = 4 \pi \dL^2(z) S[E_a,E_b],
\label{eq:ls}  
\end{equation}
where $\dL = (1+z)^2 d_\A$ is the luminosity distance. We adopt the
observed $L_\bol - T_\gas$ relation parameterized by
\begin{equation}
  L_\bol = L_{44} \left( \frac{T_{\gas}}{6{\rm keV}} 
\right)^{\alpha}
  (1+z)^\zeta ~~ 10^{44} h^{-2}{\rm ~ erg~sec}^{-1} .
\label{eq:lt}
\end{equation}
We take $L_{44}=2.9$, $\alpha=3.4$ and $\zeta=0$ as a fiducial set of
parameters on the basis of recent observational indications (David et
al. 1993; Ebeling et al. 1996; Ponman et al. 1996; Mushotzky \& Scharf
1997). Then we translate $L_\bol(T_{\gas})$ into the band-limited
luminosity $L_\band[T_{\gas},E_1,E_2]$ by properly taking account of
metal line emissions (Masai 1984) in addition to the thermal
bremsstrahlung (we fix the abundance of intracluster gas as 0.3 times
the solar value).

\begin{figure}[t]
\begin{center}
  \leavevmode\psfig{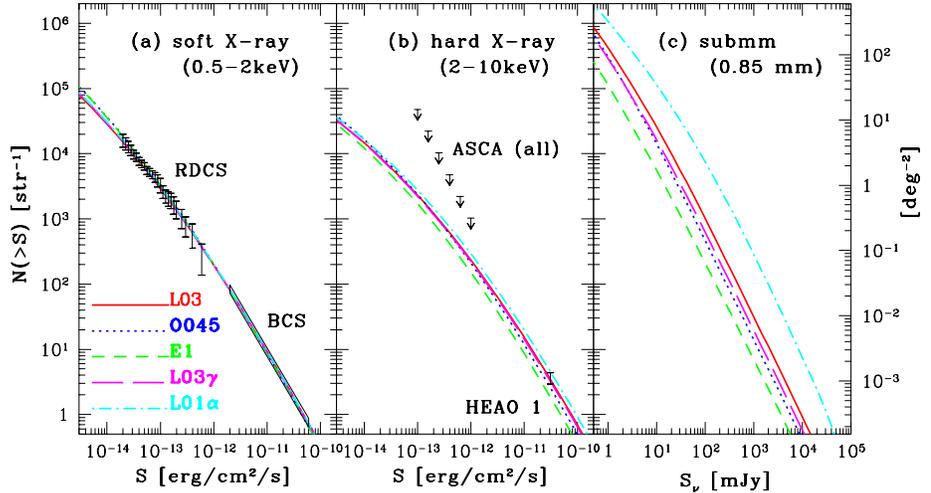}
\end{center}
\vspace*{-0.5cm}
\caption{The \ns relations of galaxy clusters for CDM models  
  in (a) the soft X-ray (0.5-2.0 keV) band, (b) the hard X-ray (2-10
  keV) band, and (c) the submm (0.85 mm) band. Lines represent the
  models listed in table 1; L03 (solid), O045 (dotted), E1 (short
  dashed), L03$\gamma$ (long dashed), and L01$\alpha$
  (dot-dashed). Also plotted in panel (a) are the 1$\sigma$ error bars
  from the RDCS (Rosati et al. 1995, 1997), and the error box from the
  BCS (Ebeling et al. 1997).  In panel (b), the arrows indicate the
  \ns of all X-ray sources in the 2-10 keV band from {\it ASCA}
  (Cagnoni et al. 1997), and the error bar (1 $\sigma$) is the number
  of clusters observed by {\it HEAO 1} (Piccinotti et al. 1982).
}
\label{fig:logns3}
\vspace*{-0.25cm}
\end{figure}

Assuming that the intracluster gas is isothermal, its temperature
$T_{\gas}$ is related to the total mass $M$ by
\begin{eqnarray}
  k_\B T_{\gas} &=& \gamma {\mu m_p G M \over 3 r_{\rm vir}(M,z_f)},
  \nonumber \\ &=& 5.2\gamma (1+z_f) \left({\Delta_{\rm vir} \over
      18\pi^2}\right)^{1/3} \left({M \over 10^{15} h^{-1} M_\odot}
  \right)^{2/3} \Omega_0^{1/3} ~{\rm keV}.
\label{eq:tm}
\end{eqnarray}
where $k_\B$ is the Boltzmann constant, $G$ is the gravitational
constant, $m_p$ is the proton mass, $\mu$ is the mean molecular weight
(we adopt $\mu=0.59$), and $\gamma$ is a fudge factor of order unity
which may be calibrated from hydrodynamical simulations or
observations.  The virial radius $r_{\rm vir}(M,z_f)$ of a cluster of
mass $M$ virialized at $z_f$ is computed from $\Delta_{\rm vir}$, the
ratio of the mean cluster density to the mean density of the universe
at that epoch. We evaluate this quantity using the formulae for the
spherical collapse model presented in Kitayama \& Suto (1996b) and
assuming for simplicity that $z_f$ is equal to the epoch $z$ at which
the cluster is observed.  Finally, we compute the mass function
$n_M(M,z)dM$ in equation (\ref{eq:logns}) using the Press--Schechter
theory (Press \& Schechter 1974) assuming $z=z_f$ as above.  The
effect of $z_f \neq z$ is discussed by Kitayama \& Suto (1997) in this
context, and the more general consideration of $z_f \neq z$ is given
in Lacey \& Cole (1993), Sasaki (1994), and Kitayama \& Suto
(1996a,b).

\section{Breaking the degeneracy}

The X-ray \ns for various Cold Dark Matter (CDM) models and the
resulting constraints on $\Omega_0$ and $\sigma_8$ are summarized in
Fig.\ref{fig:ns1chi2}.  Figure \ref{fig:ns1chi2} presents a clear
example of the degeneracy among viable cosmological models in a sense
that for a given value of $\Omega_0$, one can find a value of
$\sigma_8$ which accounts for the X-ray cluster \ns. Several examples
of such models are listed in Table 1.

\begin{table}
\vspace*{-0.5cm}
\caption{CDM models from the {\it ROSAT} X-ray
  Log $N$ -- Log $S$.}
\begin{center}
\begin{tabular}{ccccccc}
\hline\hline\\[-6pt]
Model & $\Omega_0$ &  $\lambda_0$  
&  $h$ &   $\sigma_8$ & $\alpha$ & $\gamma$\\ 
[4pt]\hline \\[-6pt]
L03 & 0.3  & 0.7 & 0.7 & 1.04 & 3.4 & 1.2  \\
O045 & 0.45 & 0   & 0.7 & 0.83 & 3.4 & 1.2  \\
E1 & 1.0  & 0   & 0.5 & 0.56 & 3.4 & 1.2 \\
L03$\gamma$ & 0.3  & 0.7 & 0.7 & 0.90 & 3.4 & 1.5 \\
L01$\alpha$ & 0.1  & 0.9 & 0.7 & 1.47 & 2.7 & 1.2 \\
\hline
\end{tabular}
\end{center}
\vspace*{-0.5cm}
\end{table}

\begin{figure}[t]
\begin{center}
  \leavevmode\psfig{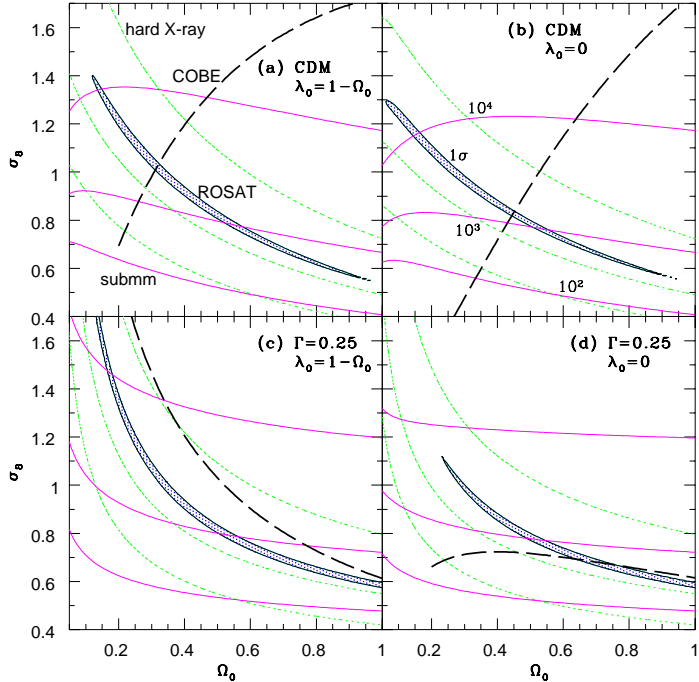}
\end{center}
\vspace*{-0.5cm}
\caption{Contour maps on the $\Omega_0$-$\sigma_8$ plane in (a) 
  spatially flat ($\lambda_0=1-\Omega_0$) CDM models, (b) open
  ($\lambda_0=0$) CDM models, (c) spatially flat CDM-like models with
  the fixed shape parameter ($\Gamma=0.25$), and (d) open CDM-like
  models with $\Gamma=0.25$.  In all cases, $h=0.7$, $\alpha=3.4$, and
  $\gamma=1.2$ are assumed. Shaded regions represent the 1$\sigma$
  significance contours derived in KS97 from the soft X-ray (0.5-2
  keV) \ns. Dotted and solid lines indicate the contours of the number
  of clusters greater than $S$ per steradian ($10^2$, $10^3$, $10^4$
  from bottom to top) with $S = 10^{-13}$ \unit in the hard X-ray
  (2-10 keV) band and with $S_\nu = 50$mJy in the submm (0.85 mm)
  band, respectively.  Thick dashed lines represent the {\it COBE} 4
  year result computed from the fitting formulae at
  $0.2<\Omega_0\leq1$ by Bunn \& White (1997).
}
\label{fig:cont_szac4}
\vspace*{-0.25cm}
\end{figure}

The degeneracy among viable cosmological models can be broken by
observing wider (i.e., increase the statistics using wide-field
surveys like SDSS, 2dF, PLANCK) and deeper (at higher redshifts) in
different bands.  (e.g., Eke, Cole, \& Frenk 1996; Barbosa, Bartlett,
Blanchard, \& Oukbir 1996; Fan, Bahcall, \& Cen 1997; Kitayama,
Sasaki, \& Suto 1998).  Figure \ref{fig:logns3} plots the cluster \ns
predictions in soft-Xray, hard X-ray, and submm (due to the
Sunyaev-Zel'dovich effect) for models which reproduce the observed \ns
in the soft-Xray band (Table 1). Figure \ref{fig:cont_szac4}
illustrates the extent to which one can break the degeneracy between
$\sigma_8$ and $\Omega_0$ in CDM models ($n=1$, $h=0.7$) using the
multi-band observational data.

\begin{figure}[t]
\begin{center}
  \leavevmode\psfig{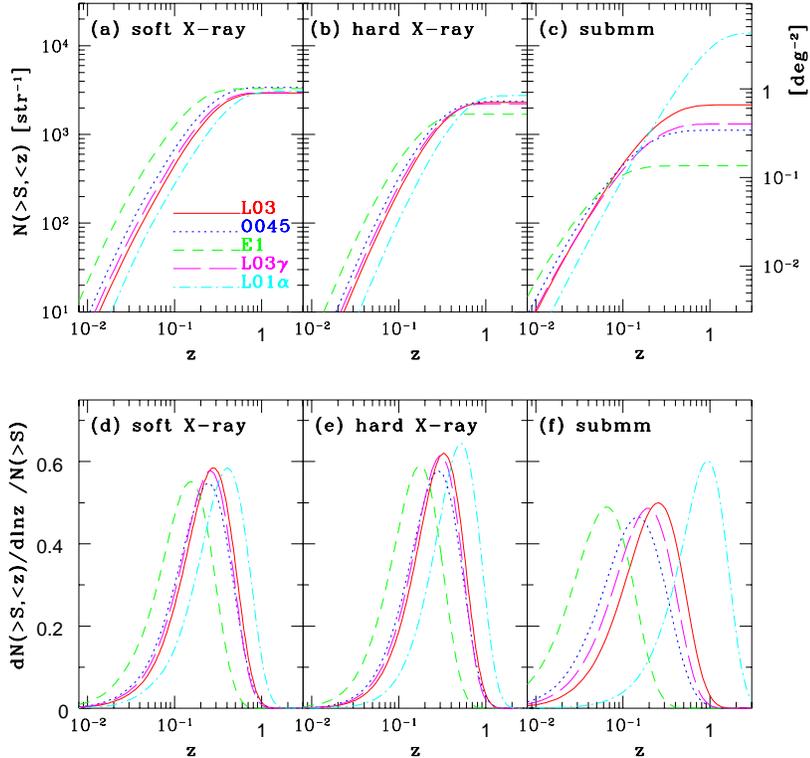}
\end{center}
\vspace*{-0.5cm}
\caption{Redshift evolution of the number of galaxy clusters. Upper 
  panels show the cumulative number $N(>\!S,<\!z)$ against $z$ in (a) the
  soft X-ray (0.5-2 keV) band with $S=10^{-13}$ \unit, (b) the hard
  X-ray (2-10 keV) band with $S=10^{-13}$ \unit, and (c) the submm
  (0.85 mm) band with $S_\nu=50$ mJy. Lower panels (d)--(f) are
  similar to (a)--(c) except for plotting the differential
  distribution $dN(>\!S,<\!z)/d\ln z$ normalized by $N(>S)$.}
\label{fig:nsz}
\vspace*{-0.25cm}
\end{figure}

\begin{figure}[h]
\begin{center}
   \leavevmode\psfig{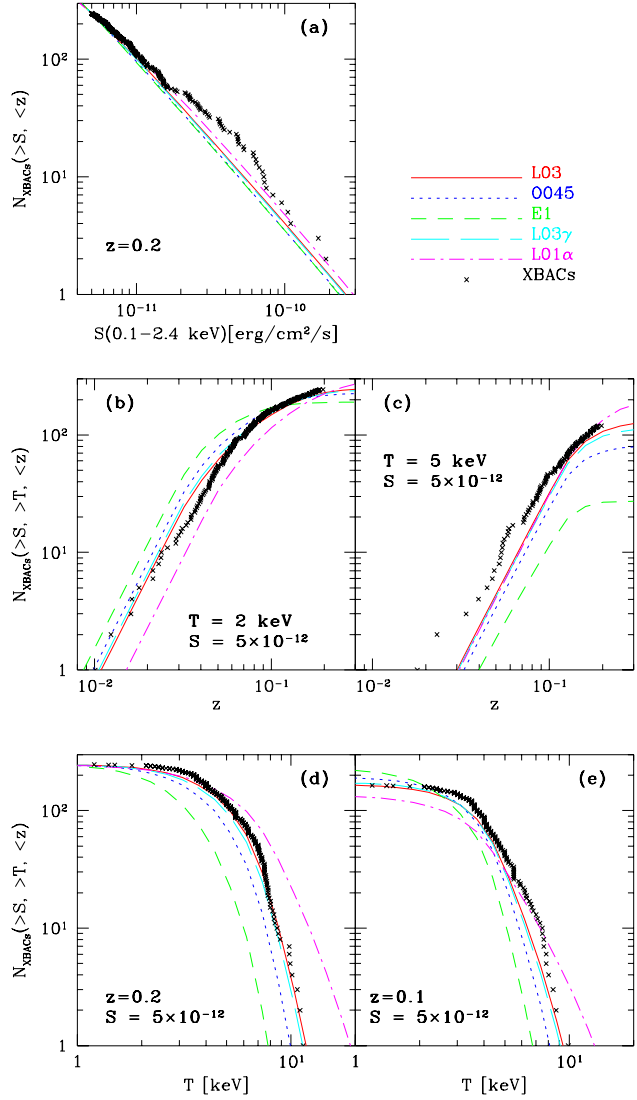}
\end{center}
\vspace*{-0.5cm}
\caption{Tentative comparison with the XBACs (Ebeling
  et al. 1996). Upper panel (a) shows the \ns in the 0.1-2.4 keV band
  with the redshift limit of $z=0.2$ (models L03 and O045 almost
  overlap with L03$\gamma$ and E1 respectively). Middle panels exhibit
  $N(>\!S,>\!T,<\!z)$ versus $z$ for $S(\mbox{0.1-2.4 keV})=5 \times
  10^{-12}$ \unit in the cases of (b) $T=2$keV and (c) $T=5$keV. Lower
  panels plot $N(>\!S,>\!T,<\!z)$ versus $T$ for $S(\mbox{0.1-2.4
    keV})=5 \times 10^{-12}$ \unit in the cases of (d) $z=0.2$ and (e)
  $z=0.1$.  All the model predictions are normalized to reproduce the
  total number of the XBACs at its flux limit $S(\mbox{0.1-2.4 keV})=5
  \times 10^{-12}$ \unit and redshift limit $z=0.2$.}
\label{fig:xbacs}
\vspace*{-5.5cm}
\end{figure}

Similarly the redshift-distribution of cluster abundances can be a
very powerful discriminator of the different cosmological
models. Figure \ref{fig:nsz} exhibits the redshift evolution of the
number of clusters in different bands. As expected, the evolutionary
behavior strongly depends on the values of $\Omega_0$ and $\sigma_8$;
the fraction of low redshift clusters becomes larger for greater
$\Omega_0$ and smaller $\sigma_8$. It is indicated that one may be
able to distinguish among these models merely by determining the
redshifts of clusters up to $z \sim 0.2$.  For this purpose, we adopt
tentatively the X-ray brightest Abell-type clusters (XBACs, Ebeling et
al. 1996), which is about 80 \% complete and consists of 242 clusters
with $S>5 \times 10^{-12}$ \unit in the 0.1-2.4 keV band and $z<0.2$.
Keeping in mind the incompleteness of the sample and uncertainties
especially in the estimated temperature data, however, we simply
intend to perform a crude comparison with our predictions in
Fig. \ref{fig:xbacs}.

The sky coverage of the XBACs is hard to quantify mainly due to the
uncertain volume incompleteness of the underlying optical catalogue as
noted by Ebeling et al. (1996).  In Fig. \ref{fig:xbacs}, therefore,
we simply plot the real numbers of the XBACs and normalize all our
model predictions to match the total number of the XBACs at its flux
limit $S(\mbox{0.1-2.4 keV})=5 \times 10^{-12}$ \unit and redshift
limit $z=0.2$.  In general, the model predictions are shown to be
capable of reproducing well the shape and amplitude of the observed
distributions $N(>\!S,>\!T,<\!z)$ even when $T$ and $z$ are varied.
Taking into account the incompleteness of the observed data and large
statistical fluctuations at low numbers, the agreements with models
L03 and L03$\gamma$ (both has $\Omega_0=0.3$) are rather
remarkable. Since the shapes and amplitudes of the predicted curves in
these figures are primarily determined by the value of $\Omega_0$,
this result provides a further indication for low $\Omega_0$ universe.

\newpage
\section{Summary}

Let us summarize the conclusions of the present talk.

\begin{description}
\item{(1)} There exist { several theoretical models} which
  successfully reproduce the observed \ns relation of galaxy clusters
  over almost four orders of magnitude in X-ray flux.

\item{(2)} The resulting $\sigma_8$ is given by the following empirical
fit (95\% confidence limit):
\begin{equation}
\sigma_8 = (0.54 \pm 0.02 \pm 0.1) \times
   \Omega_0^{-0.35-0.80\Omega_0+0.55\Omega_0^2} 
\end{equation}
for $\lambda_0=1-\Omega_0$ CDM, and
\begin{equation}
\sigma_8 = (0.54 \pm 0.02 \pm 0.1) \times
        \Omega_0^{-0.28-0.91\Omega_0+0.68\Omega_0^2} 
\end{equation}
for $\lambda_0=0$ CDM.

\item{(3)} Low-density CDM models ($n=1$) with
  $(\Omega_0,\lambda_0,h,\sigma_8) = (0.3,0.7,0.7,1)$ and $(0.45, 0,
  0.7, 0.8)$ simultaneously account for the cluster \ns, XTF, the {\it
    COBE} 4 year normalization.
\end{description}

Maybe the most important point is that many cosmological models are
more or less successful in reproducing the structure at redshift
$z\sim0$ {\it by construction}. This is because the models have still
several degrees of freedom or {\it cosmological parameters} which can
be appropriately {\it adjusted} to the observations at $z\sim0$
($\Omega_0$, $\sigma_8$, $h$, $\lambda_0$, $b(r,z)$).  We have shown
that, given a complete flux limited cluster sample with redshift
and/or temperature information, one can further constrain the
cosmological models. In fact, our tentative comparison indicates that
our predictions reproduce well the evolutionary features of the XBACs
and that the results, although preliminary, seem to favor low density
($\Omega_0 \sim 0.3$) universes.  As indicated by this preliminary
result, surveys of objects at at high redshifts in several different
bands (X-ray, radio and submm) are the most efficient and rewarding to
break the degeneracy among the viable cosmological models.

\acknowledgments We thank H. Ebeling and P. Rosati for generously
providing us their X-ray data, and K. Masai for making his X-ray code
available to us. Numerical computations presented here were carried
out on VPP300/16R and VX/4R at the Astronomical Data Analysis Center
of the National Astronomical Observatory, Japan, as well as at RESCEU
(Research Center for the Early Universe, University of Tokyo) and KEK
(National Laboratory for High Energy Physics, Japan).
T.K. acknowledges support from a JSPS (Japan Society for the Promotion
of Science) fellowship (09-7408).  This research was supported in part
by the Grants-in-Aid by the Ministry of Education, Science, Sports and
Culture of Japan (07CE2002) to RESCEU, and by the Supercomputer
Project (No.97-22) of High Energy Accelerator Research Organization
(KEK).


\clearpage

\begin{references}
\reference  
Bahcall, N.A. \& Cen, R.Y.\ 1993, ApJ 407 L49
\reference  
Barbosa, D., Bartlett, J. G., Blanchard, A., \& Oukbir, J.\ 1996, A\&A 314, 13
\reference
Bunn, E. F., \&  White, M. \ 1997, ApJ 480, 6
\reference
Cagnoni, I., Della Ceca, R., \& Maccacaro, T. \ 1997, astro-ph/9709018 
\reference
David, L. P., Slyz, A., Jones, C., Forman, W., \& Vrtilek, 
S. D. \ 1993, ApJ 412, 479
\reference
Ebeling H., et al. \ 1998, MNRAS submitted 
\reference
Eke, V. R., Cole, S., \& Frenk, C. S. \ 1996, MNRAS 282, 263
\reference
Evrard, A.E., \& Henry, J. P. \ 1991, ApJ 383, 95
\reference  
Fan, X., Bahcall, N.A. \& Cen, R.Y.\ 1997, ApJ 490 L123
\reference
Henry, J. P., \& Arnaud, K. A. \ 1991, ApJ 372, 410
\reference
Kitayama, T., Sasaki, S., \& Suto, Y.\ 1998, PASJ 50, 1
\reference
Kitayama, T., \& Suto, Y.\ 1996a, MNRAS 280, 638
\reference
Kitayama, T., \& Suto, Y.\ 1996b, ApJ 469, 480
\reference
Kitayama, T., \& Suto, Y.\ 1997, ApJ 490, 557
\reference
Lacey, C. G., \& Cole, S. \ 1993, MNRAS 262, 627
\reference
Masai, K. \ 1984, Ap\&SS 98, 367
\reference
Mushotzky, R.F., Scharf, C. A. \ 1997, ApJ 482, L13
\reference
Oukbir, J., Bartlett, J. G., \& Blanchard, A. \ 1997, A\&A 320, 365
\reference
Oukbir, J., \& Blanchard, A. \ 1997, A\&A 317, 10
\reference
Pen, U. \ 1996, astro-ph/9610147
\reference
Piccinotti, G., Mushotzky, R. F., Boldt, E. A., Holt, S. S., Marshall, 
F. E., Serlemitsos, P. J., \& Shafer, R. A. \ 1982, ApJ 253, 485   
\reference
Press, W. H., \& Schechter, P. \ 1974, ApJ 187, 425 (PS)
\reference
Ponman, T. J., Bourner, P. D. J., Ebeling, H., \& B\"{o}hringer, H. \ 
1996, MNRAS 283, 690
\reference
Rosati, P., \& Della Ceca, R. \ 1997, in preparation
\reference
Rosati, P., Della Ceca, R., Burg R., Norman, C., \& Giacconi, R. \
1995, ApJ 445, L11
\reference
Rosati, P., Della Ceca, R., Norman, C., \& Giacconi, R. \ 1997, ApJL submitted
\reference
Sasaki, S. \ 1994, PASJ 46, 427
\reference  
Shimasaku, K. \ 1993, ApJ 413 59
\reference
Sunyaev R.A., \& Zel'dovich Ya.B.\ 1972, Commts.\ Astrophys.\ Space Phys.\ 4, 
173
\reference  
Ueda, H., Itoh, M., \& Suto, Y. 1993, ApJ 408 3
\reference  
Ueda, H., Shimasaku, K., Suginohara, T., \& Suto, Y.
1994, PASJ, 46 319
\reference
Viana, P. T. P., \& Liddle, A. R. \ 1996, MNRAS 281, 323
\reference
White, S. D. M., Efstathiou, G., \& Frenk, C. S. \ 1993, MNRAS 262, 1023
\end{references}
\end{document}